\input ./style/arxiv-general.cfg
\documentclass[aoas,MSNbibl,nameyear,dvips]{arximspdf}
\makeatletter
   \@ifpackageloaded{graphicx}{}{\usepackage{graphicx}}
\makeatother

\doi{10.1214/15-AOAS873}
\volume{10}
\issue{1}
\pubyear{2016}
\firstpage{74}
\lastpage{93}
\docsubty{FLA}

\makeatletter
\newcommand{\lleft}{\left}
\newcommand{\rright}{\right}
\renewcommand{\mid}{|}
\newcommand{\eqref}[1]{(\ref{#1})}
\newcommand{\btheta}{\bolds{\theta}}
\newcommand{\given}{\mid}
\newcommand{\normal}{\mathrm{N}}
\newcommand{\bA}{\mathbf{A}}
\newcommand{\bS}{\mathbf{S}}
\newcommand{\bY}{\mathbf{Y}}
\makeatother

\begin{document}
\begin{frontmatter}

\title{A Markov-switching model for heat waves}
\runtitle{A Markov-switching model for heat waves}

\begin{aug}
\author[A]{\fnms{Benjamin A.}~\snm{Shaby}\corref{}\ead[label=e1]{bshaby@psu.edu}\thanksref{M1}},
\author[B]{\fnms{Brian J.}~\snm{Reich}\ead[label=e2]{bjreich@ncsu.edu}\thanksref{M2}},
\author[C]{\fnms{Daniel}~\snm{Cooley}\ead[label=e3]{cooleyd@stat.colostate.edu}\thanksref{M3}}
\and
\author[D]{\fnms{Cari G.}~\snm{Kaufman}\ead[label=e4]{cgk@stat.berkeley.edu}\thanksref{M4}}
\runauthor{Shaby, Reich, Cooley and Kaufman}
\affiliation{Pennsylvania State University\thanksmark{M1},
North Carolina State University\thanksmark{M2},
Colorado State University\thanksmark{M3} and
University of California, Berkeley\thanksmark{M4}}
\address[A]{B. A. Shaby\\
Department of Statistics \\
Pennsylvania State University \\
313 Thomas Building \\
University Park, Pennsylvania 16802\\
USA \\
\printead{e1}}
\address[B]{B. J. Reich\\
Department of Statistics \\
North Carolina State University\hspace*{10pt} \\
5146 SAS Hall, Box 8203 \\
Raleigh, North Carolina 27695\\
USA \\
\printead{e2}}
\address[C]{D. Cooley\\
Department of Statistics \\
Colorado State University \\
Statistics 217 \\
Ft. Collins, Colorado 80523\\
USA\\
\printead{e3}}
\address[D]{C. G. Kaufman\\
Department of Statistics \\
University of California, Berkeley \\
315 Evans Hall \\
Berkeley, California 94720\\
USA \\
\printead{e4}}
\end{aug}

%
\received{\smonth{2} \syear{2015}}
%
\revised{\smonth{8} \syear{2015}}

%
\begin{abstract}
Heat waves merit careful study because they inflict severe economic and
societal damage. We use an intuitive, informal working definition of a heat
wave---a persistent event in the tail of the temperature distribution---to
motivate an interpretable latent state extreme value model. A latent
variable with dependence in time indicates membership in the heat wave
state. The strength of the temporal dependence of the latent variable
controls the frequency and persistence of heat waves. Within each heat wave,
temperatures are modeled using extreme value distributions, with extremal
dependence across time accomplished through an extreme value Markov model.
One important virtue of interpretability is that model parameters directly
translate into quantities of interest for risk management, so that questions
like whether heat waves are becoming longer, more severe or more frequent
are easily answered by querying an appropriate fitted model. We demonstrate
the latent state model on two recent, calamitous, examples: the European
heat wave of 2003 and the Russian heat wave of 2010.
\end{abstract}

%
\end{frontmatter}

\section{Introduction}

When widespread heat waves occur, they dominate news
reports and inspire passionate discussions about climate change and public
policy. The European heat wave of 2003 was estimated
to have caused up to an estimated 70,000 additional deaths
[\citet{robine-2008a}] and cost the 2011 equivalent of \$16 billion
[\citet{munichre-2003a,ipcc-2007b}]. The Russian heat wave of
2010 was
responsible for an estimated 55,000 excess deaths, a 25\% reduction in
agriculture and \$15 billion in economic loss [\citet{barriopedro-2011a}].
Perhaps because of their high public visibility and disastrous public health
and economic consequences, heat waves are the subject of a great deal of
scientific research [e.g., \citet{easterling-2000a,huth-2000a,frich-2002a,meehl-2004a,schar-2004a,clark-2006a,fischer-2010a,otto-2012a,hanlon-2013a,amengual-2014a}]. Here
we build a model, based on an informal notion of what a heat wave is,
that may be used for studying such events. Our chief objective for
building such
a model is for it to be highly interpretable while still realistically
characterizing the upper tail of the temperature distribution.


We are aware of very few studies that have
applied extreme value theory to the analysis of heat waves. \citet
{furrer-2010a} applied a conditional points over
threshold model to daily temperatures to make inferences about the frequency,
intensity and duration of heat waves. A more recent example is
\citet{reich-2013e},
who modeled serially dependent points above a high threshold using a
transformed max-stable process. In addition to their temporal
structure, heat
waves have potentially important spatial features, which neither these works
nor ours attempt to analyze.

Part of the difficulty in analyzing heat waves might be that there is little
agreement on exactly what a heat wave is [\citet{karl-1997a,huth-2000a,palecki-2001a,khaliq-2005a}]. For example, \citet{huth-2000a} defined a heat
wave as ``the
longest continuous period (i) during which the maximum daily
temperature was
at least $T_1$ in at least three days, (ii) whose mean maximum daily
temperature was at least~$T_1$, and (iii) during which the maximum daily
temperature did not drop below $T_2$'' for some specified temperatures $T_1$
and $T_2$ [this definition was also used by \citet{meehl-2004a} and
\citet{peng-2011a}]. \citet{reich-2013e} defined a heat
wave as a run of consecutive days above some
threshold. \citet{furrer-2010a} avoided explicit definitions by
pairing their
statistical model with a stochastic weather generator, producing draws from
which the characteristics of any desired definition of a heat wave can be
inferred [the model in \citet{reich-2013e} also has this
potential]. Our model
uses an implicit definition of a heat wave according to membership in a latent
state. Once it is fit using MCMC, it can then function as a weather generator,
so it is capable of accommodating any definition of a heat wave that is
germane to a given application.

We use a
Bayesian hierarchical model with latent state variables that control whether
the temperature for each day is assigned the heat wave state or the nonheat
wave state. Temporal dependence in the latent state variables is modeled
through a simple two-state Markov chain, with one parameter in the transition
matrix controlling the frequency of heat waves and the other
controlling the
persistence of heat waves. For each day, the posterior probability of
the state
variable represents the degree of confidence with which it is
classified as
being part of a heat wave.

By employing two states, our model allows the temporal dependence of
temperatures that occur in the heat wave state to differ from the dependence
structure when temperatures are behaving ``typically.'' The heat wave
state is
modeled with a Markovian extreme-value threshold-exceedance model that allows
temperatures to exhibit extremal (asymptotic) dependence [\citet
{coles-2001a}, Section~8.4], thereby capturing the persistence of heat
waves. The nonheat wave state is modeled with Gaussian dependence
structure. Importantly,
Gaussian dependence cannot exhibit extremal dependence [\citet
{sibuya-1959a}]. A
Markov model similar to our within-heat-wave component was used by
\citet{smith-1997a}, who studied daily minimum temperature
exceedances in
Wooster, Ohio. However, whereas \citet{smith-1997a} fixed a high
threshold and
fit the Markov model to the exceedances, treating all other
observations as
censored, here we assign the extreme value Markov model to those temperatures
that are in the heat wave state, and membership in the state is
estimated from
the data.

\section{A Markov-switching model for threshold exceedances}
\label{sec:model-definition}

We begin building our model by informally defining a heat wave as a
period of
persistent extremely high temperatures. This simple notion leads
naturally to
a model with two states, one representing days that are part of a heat wave
and one representing all other days. Persistence implies positive temporal
dependence in the state variables, and extremeness implies temperatures that
lie in the upper tail of the distribution. Because our primary focus is the
behavior of the upper tail, it is important to appropriately capture tail
dependence, and we rely on models suggested by extreme value theory to
do so.
We seek a parsimonious model that represents both persistence and extremeness
in the most interpretable way possible, while still providing a
realistic fit
to the data.

To define the latent two-state model, let $S_1, \ldots, S_T \in\{0,
1\}$
denote the state of the temperature process on each day. The state variable
$S_t$ takes a value of 1 if day $t$ is in the heat wave state, and a
value of
0 otherwise. The state variables $S_1, \ldots, S_T$ are dependent in time
according to a Markov chain structure with transition matrix
\[
\bA= \lleft[\matrix{ 1-a_0 & a_0
\cr
1-a_1 & a_1} \rright].
\]
The parameter $a_0 = P(S_t=1 \given S_{t-1} = 0)$ determines the probability
of entering a heat wave, and the parameter $a_1 = P(S_t=1 \given
S_{t-1} = 1)$
determines the probability of remaining in a heat wave.

Let the time series $\bY= (Y_1, \ldots, Y_T)^{\mathrm{T}}$ denote
the observed
temperature on days $1, \ldots, T$. The distribution of each $Y_t$
will depend
on whether or not the corresponding $S_t$ positively indicates
membership in
the heat wave state. Furthermore, because daily temperature data exhibits
strong temporal dependence, we specify a dependence structure for $\bY
$, even
conditional on $\bS$. Perhaps the simplest way to model temporal
dependence in $\bY$ is through a Markov process.

\begin{figure}[b]

\includegraphics{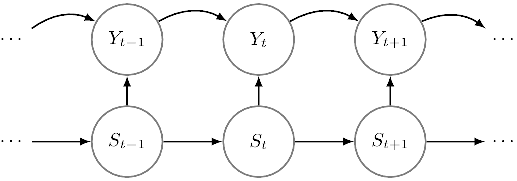}

\caption{Graphical representation of the heat wave model. The state
variables $S_1, \ldots, S_T$ are modeled as a two-state Markov
chain. The distribution of each $Y_t$ depends on its
corresponding state variable $S_t$, for $t=1, \ldots, T$. Finally,
conditionally on $S_1, \ldots, S_T$, the observations $Y_1, \ldots,
Y_T$ are also modeled as a Markov process. This structure is
sometimes referred to as a Markov-switching model. In a hidden
Markov model, there are no arrows directly connecting the
observations $Y_1, \ldots, Y_T$.}
\label{fig:model-1}
\end{figure}

Since we are assuming a Markov structure for $\bY\given\bS$, the
likelihood of $\bY\given\bS$ may be written as the product of conditional
densities
\[
L(\mathbf{y}\given\mathbf{s}) = f(y_1 \given s_1)
\prod_{t=2}^T f(y_t \given
y_{t-1}, s_t, s_{t-1}; \btheta),
\]
where $\mathbf{y}= (y_1, \ldots, y_T)^{\mathrm{T}}$ is the vector of
observed temperatures
and $\btheta$ is a vector of parameters that indexes the set of conditional
distributions. Therefore, the conditional likelihood of $\bY\given\bS
$ may be
completely specified by four families of conditional distributions $Y_t
\given
Y_{t-1}, S_t=i, S_{t-1}=j$ for $i,j \in\{0,1\}$. This type of model, depicted
graphically in Figure~\ref{fig:model-1}, is sometimes referred to as a
Markov-switching model [\citet{fruhwirth-schnatter-2006a}].
Markov-switching
models resemble hidden Markov models, the latter differing in that $Y_1,
\ldots, Y_T$ are conditionally independent given $S_1, \ldots, S_T$.

%

Conditioning on the state variables to separate the likelihood into two main
components, one arising from the heat wave state and one arising from the
nonheat wave state, endows each component with an immediate
interpretation: a
dedicated tail model for the heat wave state and a model for the bulk
of the
distribution for the nonheat wave state. Building a separate model for the
tail of the distribution is common practice in extreme value analysis.
The key
concern is that any distributional assumptions designed to fit the bulk
of the
distribution well may be insufficiently flexible to accommodate the behavior
of the tail, and attempts to fit the entire distribution, including the tail,
nonparametrically are frustrated by the dearth of data in the tail. A related
consideration when assuming the entire distribution comes from a single
parametric model is that any fitting procedure will encourage fidelity
to the
main part of the distribution at the expense of the tail for the simple reason
that, by definition, there is much more data in the bulk than in the tail.
This is especially undesirable when one is primarily interested in learning
about the tail, as we are here. The state variables provide a convenient
construct for building separate models for the tail and the main part
of the
distribution into the likelihood.

We now turn our attention to the conditional likelihoods $f(y_t \given
y_{t-1}, s_t,\break s_{t-1}; \btheta)$. In the heat wave state, we assume
that the
temperature is in the far right tail of the distribution. Extreme value theory
says that the marginal distribution of values in the upper tail is well
approximated by a generalized Pareto distribution (GPD) [\citet
{coles-2001a}],
which has survivor function
\[
P(Y > y | Y > u) = \biggl(1+\frac{\xi}{\sigma}(y-u) \biggr)_+^{-1/\xi},
\]
where $u$ is a high threshold, $\sigma$ is a scale parameter, and $\xi
$ is a
shape parameter that controls the thickness of the tail. Thus, conditionally
on $S_t=1$, we want $Y_t$ to follow a GPD. Furthermore, we want consecutive
observations $Y_{t-1}$ and $Y_t$, given $S_{t-1}=1$ and $S_t=1$ (i.e., given
both are in the heat wave state), to exhibit \emph{extremal dependence}
because it is clearly seen in the data (see Figures~\ref{fig:lag-1-frechet-scale},
\ref{fig:chi}). Extremal dependence between $Y_{t-1}$ and $Y_t$ exists if
there is positive probability that both observations lie in the asymptotic
tail of their bivariate distribution, that is, if $\lim_{c \rightarrow
\infty}P(Y_{t} > c \given Y_{t-1} > c) > 0$ for $Y_{t-1}$ and $Y_t$
having the same
marginal distribution.

To build the conditional likelihoods for the case where both $Y_{t-1}$ and
$Y_t$ are in the heat wave state, $f(y_t \given y_{t-1}, s_t=1, s_{t-1}=1;
\btheta)$, with the desired extreme value properties, we follow
\citet{smith-1997a} and construct
\begin{equation}
\label{eqn:conditional-from-joint} f(y_t \given y_{t-1}, s_{t-1}=1,
s_{t}=1; \btheta) = \frac{f(y_{t-1}, y_{t} \given s_{t-1}=1, s_{t}=1; \btheta)}{
f(y_{t-1} \given s_{t-1}=1; \btheta)},
\end{equation}
where the joint density $f(y_{t-1}, y_{t} \given s_{t-1}=1, s_{t}=1;
\btheta)$
is a parametric family with GPD margins and extremal dependence, and
$f(y_{t-1} \given s_{t-1}=1; \btheta)$ is the density of the GPD. This
definition affords some flexibility in that any valid joint density
with GPD
margins and extremal dependence may be used, and several choices for bivariate
parametric families are known [\citet{coles-2001a}, Chapter~8].
Here, we choose the
simplest such family, the logistic family with parameter $\alpha\in(0,1]$.
The bivariate logistic model may be defined through its cumulative
distribution function (CDF),
\begin{equation}
\label{eqn:logistic-copula} G(z_{t-1}, z_{t}) = \exp \bigl\{-
\bigl(z_{t-1}^{-1/\alpha} + z_{t}^{-1/\alpha}
\bigr)^\alpha \bigr\},
\end{equation}
where $z_{t-1}$ and $z_{t}$ are derived from $y_{t-1}$ and $y_{t}$ by applying
the transformation from GPD to unit Fr\'echet, $z =
-\log(F_{\mathrm{GPD}}(y))^{-1}$, where $F_\mathrm{GPD}$ denotes the
CDF of the
GPD. The bivariate likelihoods $f(y_{t-1}, y_{t} \given s_{t-1}=1, s_{t}=1;
\btheta)$ are obtained by differentiating \eqref{eqn:logistic-copula} with
respect to $y_{t-1}$ and $y_t$. For the logistic model, smaller values of
$\alpha$ indicate stronger dependence, with $\alpha\rightarrow0$
representing complete dependence and $\alpha= 1$ representing complete
independence.

The next case that we consider is when $S_{t-1}=S_{t}=0$, indicating that
times $t-1$ and $t$ are both members of the nonheat wave state. This
case is modeled
simply as an AR(1) process with mean $\mu$, variance $\sigma^2_N$ and
autocorrelation parameter $\phi\in(0,1)$ (negative
autocorrelation is physically implausible). That is, the conditional
densities for days outside of heat
waves are modeled directly as
\[
f(y_{t} \given y_{t-1}, s_{t}=0,
s_{t-1}=0) = \normal \bigl(\mu+ \phi(y_{t-1}-\mu),
\sigma_N^2 \bigr),
\]
where, unlike in the case of the logistic family, larger values of the
dependence parameter $\phi$ indicate stronger dependence. The Gaussian AR(1)
process is appropriate for the bulk of the temperature distribution (see
Figures~\ref{fig:lag-1-original-scale}, \ref{fig:pacf}), but, unlike the
logistic Markov process defined by \eqref{eqn:conditional-from-joint} and
\eqref{eqn:logistic-copula}, the AR(1) process is asymptotically independent
and therefore inadequate as a model for the tail behavior. More elaborate
models for the bulk of the temperature distribution are possible, but because
our focus is on the tail, we use the simplest available structure that seems
to fit the data, the Gaussian AR(1).

Finally, we must specify the heterogeneous cases \{$S_{t-1}=0,
S_{t}=1$\} and
\{$S_{t-1}=1, S_{t}=0$\}. These represent the transitions into and out
of heat
waves. The approach we take here is again similar to that of
\citet{smith-1997a} in that we define the conditional densities through
corresponding bivariate densities as in \eqref{eqn:conditional-from-joint},
which again have logistic dependence defined through
\eqref{eqn:logistic-copula}. The difference from the $\{S_{t-1}=S_t=1\}$
case is that here one of the marginal distributions is Gaussian rather than
GPD. This necessitates two modifications. The first is that for $\{S_{t-1}=0,
S_t=1\}$ (the transition into a heat wave), the density in the
denominator of
\eqref{eqn:conditional-from-joint} is normal. The second is that in both
heterogeneous cases, one of the $z$ variables in \eqref{eqn:logistic-copula}
is the result of a transformation from normal to unit Fr\'echet, $z =
-\log(\Phi[(y-\mu)/\sigma_N])^{-1}$, rather than
both being the result of the transformation from GPD to unit
Fr\'echet. 
Here, we use a single dependence parameter $\alpha_{01}$ to
characterize the
temporal dependence between the first day of a heat wave and the day before,
and between the last day of a heat wave and the day after.

Explicit formulas for the bivariate likelihoods for the days
corresponding to
transitions into and out of heat waves, that is, for $y_{t-1}, y_t
\given
s_{t-1}=0, s_{t}=1$ and $y_{t-1}, y_t \given s_{t-1}=1, s_{t}=0$, are
constructed as follows. First, transform both margins to $U(0,1)$ by taking,
for $j=t-1, t$,
\[
u_j = \cases{ \displaystyle\Phi \biggl(\frac{y_j-\mu}{\sigma^2_N} \biggr), &
\quad when $s_j=0$,
\cr
\displaystyle1 - \biggl[1 + \frac{\xi(y_j-u)}{\sigma}
\biggr]^{-1/\xi
}, &\quad when $s_j=1$,}
\]
where $\Phi(\cdot)$ is the standard normal CDF. Next, transform both
margins to unit Fr\'echet using $z_j = -\log(u_j)^{-1}$, for $j=t-1,
t$. The
bivariate likelihood is then
\[
f(y_{t-1}, y_t \given s_{t-1}, s_t;
\btheta) = K_{t-1}K_{t}(V_{t-1}V_{t}-V_{t-1,t})e^{V},
\]
where $V = (z_{t-1}^{-1/\alpha_{01}} + z_{t}^{-1/\alpha
_{01}})^{\alpha_{01}}$,
$V_{t-1,t} = $
$(1-1/\alpha_{01})(z_{t-1}z_{t})^{-1/\alpha_{01}-1}V^{1-2/\alpha
_{01}}$ and,
for $j=t-1, t$, $V_j = z_j^{-1/\alpha_{01}-1}V^{1-1/\alpha_{01}}$ and
\[
K_j = \cases{ \displaystyle\varphi \biggl(\frac{y_j-\mu}{\sigma^2_N} \biggr)
z_j^2 \exp(1/z_j), &\quad when
$s_j=0$,
\vspace*{3pt}\cr
\sigma^{-1}u_j^{1+\xi}z_j^2
\exp(1/z_j), &\quad when $s_j=1$,}
\]
where $\varphi(\cdot)$ is the standard normal probability density
function (p.d.f.).

The bivariate likelihoods for days within heat waves, that is, for
$y_{t-1}$, $y_t \given
s_{t-1}=1$, $s_{t}=1$, are exactly the same, only the dependence parameter
$\alpha$ is substituted for $\alpha_{01}$.

Most of the model
parameters have direct physical interpretations, so posterior inference on
them immediately tells us something about the nature of the observed heat
waves. First and foremost, the state variables $S_1, \ldots, S_N$ indicate
whether or not each day is classified as being in a heat wave. By
looking at
the posterior state probabilities for each day, we can easily retrospectively
identify when and for how long heat waves occurred, according to the
model. The Markov transition
probability $a_0$ represents the propensity of the system to enter into heat
waves, and $a_1$ represents the propensity of heat waves to persist
once they
get started. Hence, together $a_0$ and $a_1$ describe the expected
number and
duration of heat waves. The GPD parameters $u, \sigma$ and $\xi$ characterize
the severity of the heat waves, with $u$ representing the minimum temperature
needed to attain heat wave status. The dependence parameters $\alpha$,
$\alpha_{01}$ and $\phi$ together control the strength of the temporal
dependence in the temperature series. Finally, $\mu$ and $\sigma_N^2$ describe
the marginal behavior of the temperature on days that are not in heat waves.

An interesting feature of the model is that short-lived extremely high
temperatures are not necessarily classified as heat waves. The Markov chain
structure of the state variables encourages the model to consider
duration in
its classification criteria, so single very hot days, for example, will
tend to
have low posterior probability of being in the heat wave state. This
phenomenon is illustrated in the results of the case study found in the next
section.

\section{Case studies}
\label{sec:case-study}

As case studies, we select temperature time series from Paris and Moscow,
which both recently suffered through high-profile heat waves. To
simplify the
analysis, we extract daily maximum temperatures from the summer months
(JJA) from
the years 1990--2011 (92 days per year over 22 years), a time period
that is
short enough that a stationarity assumption is plausible. Temperature
data is
available through the European Climate Assessment Database
(\surl{http://eca.knmi.nl/}). For the time period under consideration, the
Paris time series is complete, and the Moscow time series contains just two
missing values. To remove seasonal effects in the JJA data, we de-seasonalize
using the following procedure. First, we fit a penalized spline to the JJA
temperatures, using absolute error, rather than squared error, as a loss
function (using the \texttt{qsreg} function in the \texttt{R} package
\texttt{fields} [\citet{R-fields}], with the smoothing parameter
chosen by the
default generalized cross-validation criterion). In this way, we do not allow
the magnitude of the extremes to unduly influence the calculation of the
climatological average. Next, we subtract the fitted spline function
from the
raw data. Finally, to aid interpretation, we add back the (constant in time)
overall median temperature so that the magnitude remains on the same
scale as
the original data, but without the seasonal cycle. Even after removing the
seasonally-varying median in this way, it is possible that some seasonality
remains, for example, a seasonally-varying variance, although we find no
evidence of this. With the (first-order) seasonal cycle removed, now assume
that all model parameters are constant in time.

\subsection{Exploratory analysis}

\begin{figure}

\includegraphics{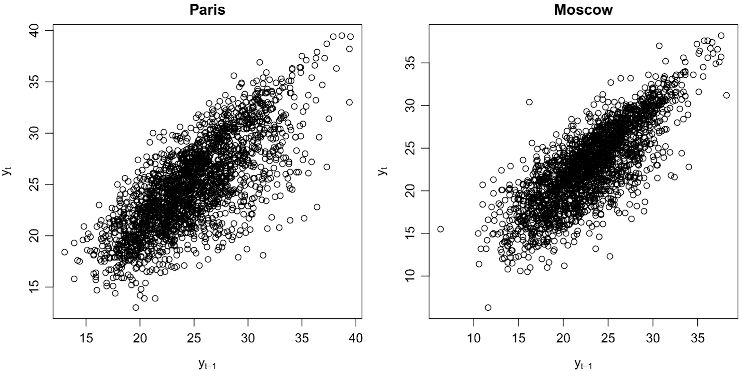}

\caption{JJA temperatures $y_t$ (i.e., at day $t$) plotted against the
temperature
of the previous day, $y_{t-1}$ (in $^\circ$C). Paris is shown in the
left panel,
Moscow on the right panel. There is strong autocorrelation in both
cases.}
\label{fig:lag-1-original-scale}
\end{figure}

To check the validity of the model, we run a variety of diagnostics.
Since most of the data does not lie in heat waves, the AR(1) portion of
the model is the easiest to check. A scatterplot of $y_{t-1}$ vs. $y_t$ is
shown in Figure~\ref{fig:lag-1-original-scale}, which indicates strong
autocorrelation at lag 1 in both Paris and Moscow.

%

\begin{figure}[b]

\includegraphics{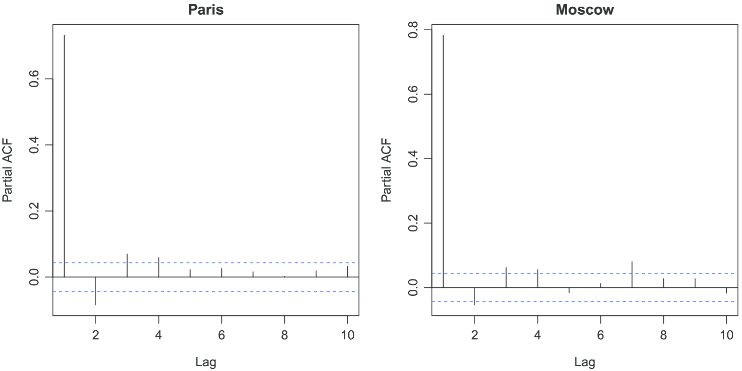}

\caption{Partial autocorrelation functions for the temperature data. Paris
is shown in the left panel, Moscow on the right panel. The large
value at lag 1 and the small values at all other lags are consistent
with the AR(1) assumption. Values between the dashed lines are not
significantly different from zero at $\alpha=0.05$.}
\label{fig:pacf}
\end{figure}

Next, we plot empirical partial autocorrelation functions for each city
in Figure~\ref{fig:pacf}. In both cities, we see a large value at lag
1, quickly decaying
to near zero by lag 2. This pattern is consistent with an AR(1) model. To
further check the validity of the AR(1) assumption, we fit AR($p$)
models to
each year of data separately and choose $p$ using AIC. In the majority of
years, AIC chooses $p=1$. We conclude that the simple structure we have
specified for the nonheat wave days is adequate.

\begin{figure}[t]

\includegraphics{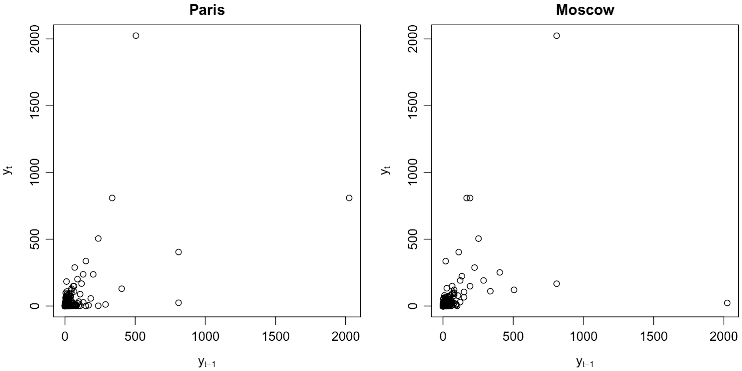}

\caption{JJA daily temperatures $y_t$ plotted against the temperature
of previous day, $y_{t-1}$, on the Fr\'echet scale. Paris is shown in
the left panel,
Moscow on the right panel. The presence of many points lying in the
interior of the plot (i.e., away from the axes) suggests strong asymptotic
dependence at lag 1.}
\label{fig:lag-1-frechet-scale}
\end{figure}

To check for extremal dependence, we again look at a scatterplot of $y_{t}$
against $y_{t-1}$, but this time we first transform the data to the
Fr\'echet
scale using a rank transformation (Figure~\ref{fig:lag-1-frechet-scale}). If
asymptotic dependence were not present, Figure~\ref{fig:lag-1-frechet-scale}
would show points lining up along the $y_{t-1}$- and $y_t$-axes. What
we see
instead is that points lie in the interior of the plot, a pattern that
indicates asymptotic dependence [\citet{coles-2001a}].

%

%

%
\begin{figure}[b]

\includegraphics{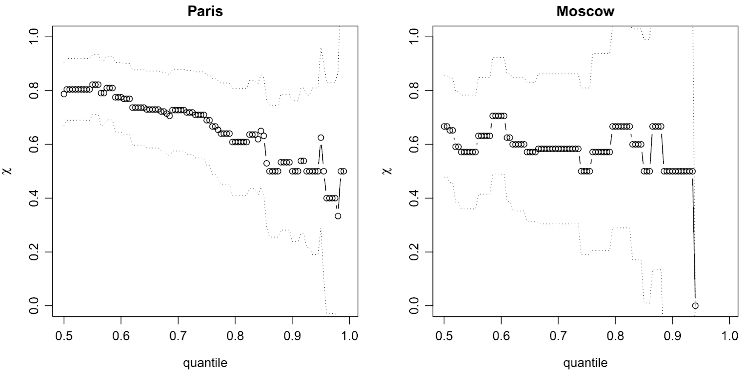}

\caption{Estimates of $\chi(u)$, for increasing values of $u$ (with the
$u$-axes on the quantile scale). Paris is shown in the left panel,
Moscow on the right panel. Since $\chi=0$ indicates extremal
independence, these plots indicate the presence of asymptotic
dependence.}
\label{fig:chi}
\end{figure}

As an additional check for extremal dependence, we examine estimates of the
quantity $\chi= \lim_{u \rightarrow\infty} \chi(u)$, where $\chi
(u) = P\{Y_t > u
\given Y_{t-1} > u\}$ [\citet{coles-1999a}], for the two cities.
A value of $\chi=0$
indicates asymptotic independence, while any $0<\chi<1$ indicates asymptotic
dependence. In practice, we estimate $\chi(u)$ for many values of $u$ and
examine its behavior as $u$ gets large. Plots of $\hat{\chi}(u)$ are
shown in
Figure~\ref{fig:chi}, with the $x$-axis transformed to the quantile
scale for
clarity. In both cases, the curves remain comfortably away from zero, except
at the far right-hand edge of the Moscow plot where there is almost no data,
again suggesting that asymptotic dependence is present in the data at
lag 1.
(For additional exploratory analysis, see the Supplemental Materials
[\citet{shaby-2015a-supp}].)

\subsection{Prior specification and computing}
Since these exploratory checks are consistent with the proposed model,
we move
ahead. The next step is to specify prior distributions on the model
parameters. For the GPD marginal parameters, we choose a vague normal for
$\log\sigma$, a uniform on $(-0.5, 0.5)$ for $\xi$, and a normal
with a small
variance centered on the $0.98$ quantile (33$^\circ$  for Moscow and
35$^\circ$
for Paris) for the threshold $u$. The priors on $\xi$ and $u$ are informative,
but, we believe, justified. Previous studies of summer high temperatures
routinely estimate $\xi$ at around $-0.22$, so the chosen uniform
prior will
have little effect other than ensuring that the posterior have no
support on
$(-\infty, -0.5]$, the region for which the GPD is not regular (i.e., standard
likelihood results do not apply [\citet{smith-1985a}]). The tight
normal prior
on $u$ encourages the GPD to be applied only to the tail of the distribution,
but not so far into the tail as to be irrelevant. The informative prior
on $u$
is necessary to achieve good convergence, and less restrictive than the
standard practice in extreme value analysis of fixing $u$ at a prespecified
value. A sensitivity analysis for the prior on $u$ is reported in the
Supplementary Materials.

For the Gaussian marginal parameters, we choose vague
normal priors for both $\mu$ and $\log\sigma_N^2$. The dependence parameters
$\alpha$, $\alpha_{01}$ and $\phi$ are given uninformative uniform $(0,1)$
priors. Finally, conjugate beta priors are specified for the Markov transition
probabilities $a_0$ and $a_1$.

Posterior simulation is carried out using a block Gibbs sampler, with
conjugate updates for $a_0$ and $a_1$, and Metropolis updates for all other
model parameters. The state variables $S_1, \ldots, S_T$ are updated jointly
using a forward-filtering backward-sampling algorithm
[\citet{fruhwirth-schnatter-2006a}] within the Gibbs sampler.
Missing values are
handled seamlessly by treating them as unknown parameters and drawing from
their predictive distributions at each MCMC iteration.

\subsection{Results}

\begin{figure}[t]

\includegraphics{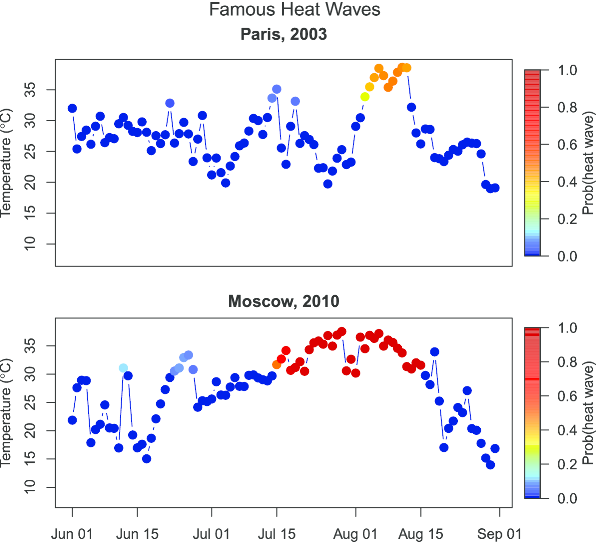}

\caption{Temperature data from the European heat wave of 2003, which hit
France especially hard, and the Russian heat wave of 2010. Color
coding of the dots indicates the posterior probability of being in
State 1. Note that in Moscow, August 17 was hotter than, say, August 14,
but the model did not classify it as a heat wave because it was a
single hot day. Similarly in Paris in the middle of July.}
\label{fig:famous-heatwaves}
\end{figure}

For each day in the study period, the sampler outputs the posterior
probability of being in a heat wave. A useful place to start examining the
results is by looking at these posterior probabilities for time periods that
include the famous heat waves that motivated this study. Figure~\ref{fig:famous-heatwaves} shows the temperature time series from the summer
of 2003 in Paris and the summer of 2010 in Moscow.

%

The $y$-axis in Figure~\ref{fig:famous-heatwaves} is the observed temperature,
and the color of each dot is proportional to the posterior probability of
membership in the heat wave state. These plots show that the model correctly
identifies these well-known events. Furthermore, it locates fairly clear
beginning and end points of each event, the locations of which are not so
obvious from just looking at the time series in the case of Moscow in 2010.

The bottom panel of Figure~\ref{fig:famous-heatwaves} also
demonstrates the
interesting feature described at the end of Section~\ref{sec:model-definition}, where a very hot day in Moscow on August 17 is not
classified as a heat wave, even though it was noticeably warmer than
other days that
were classified as heat waves. This is because the value of $a_1$ that the
model estimates defines a heat wave as having rather strong persistence with
high probability, and August 17 stands apart from its closest neighbors,
whereas the cooler days toward the end of the 2010 heat wave were
members of a
contiguous mass. This feature of the model conforms to our notion of
what a
heat wave is: it must be both very hot and persistent---just being hot
is not
enough. In this way, the fitted model contains an implicit definition
of a
heat wave, inclusion in State 1 given the data.\looseness=-1

%
\begin{figure}[t]

\includegraphics{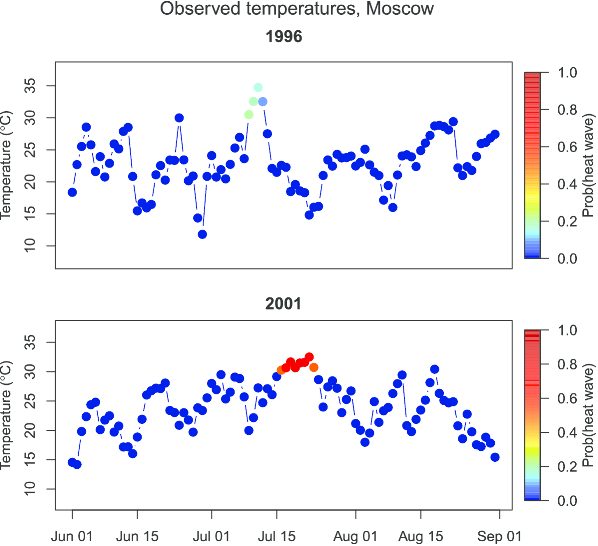}

\caption{Even though the annual maximum in 1996 was higher than that
in 2001, the model classifies the hot period in 2001 more
confidently as a heat wave, probably because of the prolonged
period and strong temporal dependence in 2001.}
\label{fig:two-moscow-heatwaves}
\end{figure}

Figure~\ref{fig:two-moscow-heatwaves} shows two summer high
temperature time
series from Moscow, 1996 and 2001. The model output suggests that there might
have been a heat wave in 1996. However, even though the annual maximum
in 1996
was higher than the annual maximum in 2001, the hot period in 2001 was
classified more confidently as being a heat wave. This is again because of
persistence; the hot period in 1996 lasted a short time and showed weak
temporal dependence, while the hot period in 2001, though cooler, lasted
longer and showed stronger temporal dependence consistent with the posterior
estimate of $\alpha$.

\begin{figure}[t]

\includegraphics{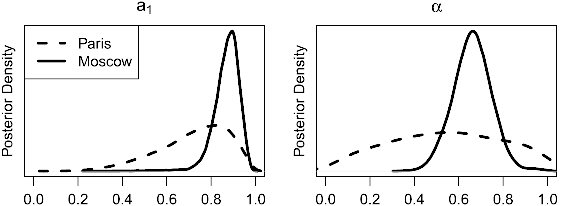}

\caption{Kernel density estimates of the marginal posterior densities of
$a_1$ and $\alpha$.}
\label{fig:posterior-dependence}
\end{figure}

\begin{figure}

\includegraphics{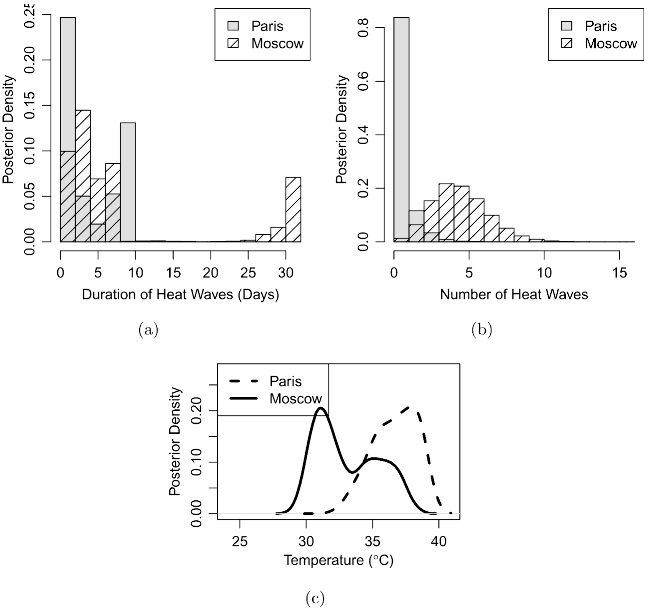}

\caption{A retrospective analysis of heat waves during the study period.
Panel \textup{(a)} shows the
posterior p.m.f. of the length of the heat waves that occurred
in Paris (shaded) and Moscow (crosshatched), and panel
\textup{(b)} shows the posterior
p.m.f. of the number of heat waves that occurred. Panel
\textup{(c)} shows that the
temperatures that occurred during heat waves were much higher in
Paris than in Moscow.}
\label{fig:retrospective-histograms}\label
{fig:retrospective-histogram-length}\label
{fig:retrospective-histogram-number}\label
{fig:retrospective-density-temperature}
\end{figure}

Figure~\ref{fig:posterior-dependence} shows kernel density estimates
of the
posterior densities of the parameters $a_1$ and $\alpha$. The Markov
transition probability $a_1$ is the probability of remaining in a heat wave,
given that one has already started. Comparing the two curves, it
appears that
the model is more confident that Moscow (solid curve) tends to have persistent
heat waves than it is about Paris (dashed curve), although for both cities
almost all of the mass lies well to the right of 0.5. The logistic
dependence parameter $\alpha$ controls the temporal dependence of temperatures
within heat waves. From Figure~\ref{fig:posterior-dependence}, we see that
while posterior means for the two cities are similar, the model again allows
posterior mass to concentrate more for Moscow at around 0.6, well away from
the extreme cases of independence and complete dependence. The
relatively high posterior precision in Moscow probably reflects the larger
number of heat waves that occurred there during the study period
[Figure~\ref{fig:retrospective-histogram-number}(b)]. The combined
interpretation of the
two parameters shown in Figure~\ref{fig:posterior-dependence} is that with
high probability, Moscow, when it does experience heat waves, tends to
experience longer heat waves with more stable temperatures.

The expected length and frequency of the heat waves is directly
calculable from
$a_0$ and $a_1$, and posterior distributions of these expectations are
straightforward to estimate from the MCMC sample. This type of exercise
is useful
for making predictions in a stationary world. In addition to looking ahead
using expectation-type calculations, it is interesting to do retrospective
analysis of actual heat waves during the study period.

%

%

Figure~\ref{fig:retrospective-histograms} demonstrates a retrospective
analysis. The left-hand panel (a)
shows the posterior probability mass function (p.m.f.) of the length of
heat waves that occurred in
Paris and Moscow from 1990--2011. The most prominent feature of
Figure~\ref{fig:retrospective-histogram-length}(a)
is the large amount of probability
mass for Moscow at large durations. This massive right tail mostly reflects
the extremely long 2010 heat wave. Figure~\ref{fig:retrospective-histogram-number}(b) shows that heat waves in Moscow
tended to be more numerous than those in Paris. However, we see in
Figure~\ref{fig:retrospective-density-temperature}(c) that heat waves
in Paris tended to be
much hotter. Putting together these three characteristics, it appears that
heat waves in Paris from 1990--2011 were hotter, though shorter and less
frequent, than those in Moscow.

An anonymous referee points out that this behavior is as expected.
Moscow has
a more continental climate, enabling stable anticyclonic conditions (or
\emph{blocking} episodes), associated with clear skies and excesses of
downward solar radiation, to persist for long periods. In contrast, Western
Europe is under the influence of the jet stream and its westerly winds
directly coming from the Western Atlantic ocean, which inhibits the
maintenance of blocking situations. The higher temperatures observed in
Paris are likely due to the effect of latitude.

\subsection{Alternative definitions of heat waves}
\label{sec:weather-generator}

To explore the behavior of the model under alternative definitions of heat
waves, we use it as a stochastic weather generator and compute the posterior
distribution of frequency, duration and mean temperature of heat waves, where
heat waves are defined according to criteria found in the literature.
For each
MCMC iteration, we simulate 500 summers worth of random draws from the model,
conditional on the model parameters at that iteration. This results in a
posterior sample of summers, from which we can apply any definitions of heat
waves that we choose. Following \citet{meehl-2004a}, we use two common
criteria. The first defines a heat wave as the three-day period in any
given year with the
highest average low temperature, based on the idea that stretches without
relief from extreme heat may have large health impacts [\citet
{karl-1997a}].
Since we are working with daily high temperatures rather than daily
lows, we
modify this ``worst annual event'' definition accordingly. The second considers
two thresholds $T_1$ and $T_2$ and defines a heat wave as the longest
contiguous period during which the daily high temperature exceeds $T_1$ at
least three times, the daily high temperature is always above $T_2$,
and the
average daily high temperature is greater than $T_1$ [\citet
{huth-2000a}]. The
thresholds $T_1$ and $T_2$ are set, respectively, at the 0.975 and 0.81
empirical quantiles [\citet{meehl-2004a}].

%
\begin{figure}[b]

\includegraphics{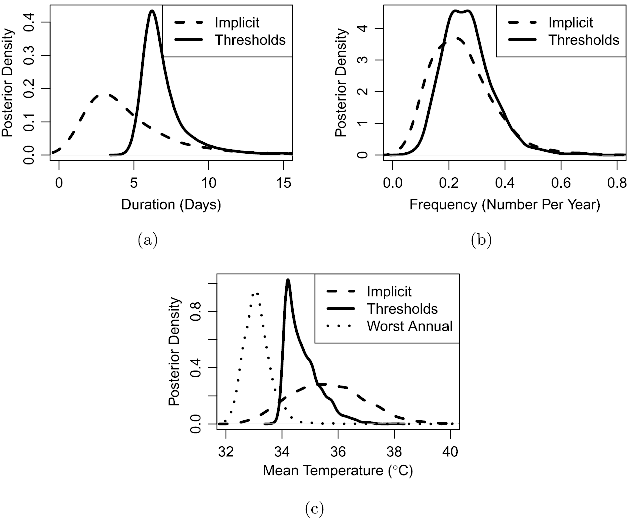}

\caption{Comparisons of Paris heat waves under different definitions. Panel
\textup{(a)} shows kernel density estimates of
the posterior distribution of the duration of heat waves, where
heat waves are defined implicitly by the latent state model, as
well as using a threshold definition. The threshold definition
produces heat waves that are longer than the implicit definition.
Panel \textup{(b)} shows the posterior
densities of the frequency of heat waves under the same two
definitions of heat waves. The two definitions (coincidentally)
produce heat waves at similar frequencies. Panel
\textup{(c)} shows the posterior densities of the
mean daily high temperatures during heat waves, under the two
previous definitions, plus the ``worst annual event'' definition.
This latter definition is less restrictive, and hence produces heat
waves that are cooler than the other two. All posterior
distributions are sampled by drawing from the latent state model,
conditional on model parameters at each iteration of the MCMC
sampler.}
\label{fig:paris-weather-gen}\label{fig:paris-gen-duration}\label
{fig:paris-gen-frequency}\label{fig:paris-gen-temp}
\end{figure}

\begin{figure}

\includegraphics{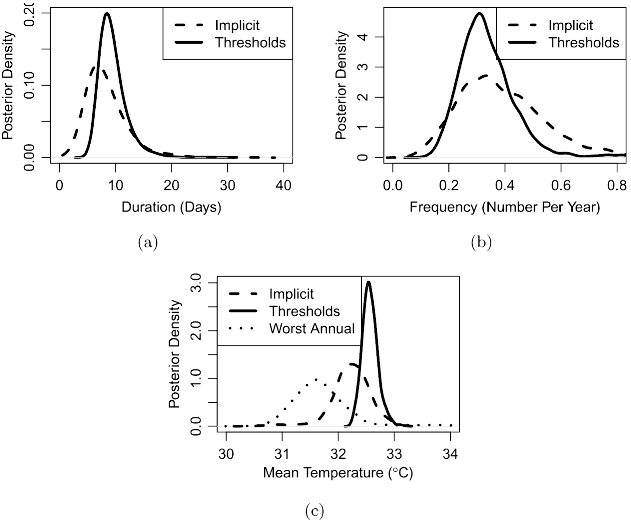}

\caption{Comparisons of Moscow heat waves under different definitions.
Panels \textup{(a)},
\textup{(b)} and \textup{(c)}
show the same basic patterns seen in Figure~\protect\ref
{fig:paris-weather-gen}. In comparison with Figure~\protect\ref
{fig:paris-weather-gen}, the most noticeable difference is that
posterior densities for the implicit definition and the threshold
definition coincide more closely for Moscow than for Paris. Just as
for Paris, Panel \textup{(c)} shows that the
``worst annual event'' definition produces the least severe heat
waves, as expected.}
\label{fig:moscow-weather-gen}\label{fig:moscow-gen-duration}\label
{fig:moscow-gen-frequency}\label{fig:moscow-gen-temp}
\end{figure}

Figures~\ref{fig:paris-weather-gen} (Paris) and \ref{fig:moscow-weather-gen}
(Moscow) show features of our simulated heat waves, under the three
definitions of heat waves (the definition implicit in our latent state
model---a contiguous block of days for which $S=1$, the threshold-based
definition, and the ``worst annual event'' definition). We plot the posterior
density of the duration of heat waves [Figures~\ref{fig:paris-gen-duration}(a)
and \ref{fig:moscow-gen-duration}(a)], the frequency of heat waves
[Figures~\ref{fig:paris-gen-frequency}(b) and \ref
{fig:moscow-gen-frequency}(b)] and the
mean temperature during heat waves [Figures~\ref{fig:paris-gen-temp}(c) and
\ref{fig:moscow-gen-temp}(c)]. By definition, the ``worst annual
event'' type of
heat wave occurs exactly once per year for three days, making duration and
frequency trivial.

The overall patterns in heat wave characteristics are similar across
cities. The implicit latent state definition produces shorter heat
waves than
the threshold definition. The frequency of implicitly-defined and
threshold-based heat waves is similar, but the posterior distribution is
slightly more diffuse for the implicit definition. For the mean daily high
temperature during heat waves, the ``worst annual event'' heat waves are
cooler than the other two, which is expected because the implicit and
threshold definitions find heat waves less frequently than once
annually, and
hence exclude the less extreme annual events. The posterior
distribution of
mean temperatures is noticeably more peaked for the threshold than for the
implicit definition.

%

\subsection{Assessing model fit}
\label{sec:model-fit}

Popular tools for assessing the fit of a Bayesian model to a given data set
(e.g., deviance information criterion [\citet
{spiegelhalter-2002a}], proper
scoring rules [\citet{gneiting-2007a}]) work by comparing the
fits of competing
models and choosing the one with the highest score. Here, our goal is
interpretability, not achieving the best possible fit to the data, so these
tools are not ideal. However, we still need to determine whether the
fit is adequate. To check compatibility with the data without making
comparisons among competing models, we use the posterior predictive
checks of
\citet{gelman-1996a}. The idea is to make posterior predictive
draws from the
model and see whether those draws resemble the observed data set
according to
a suite of relevant summary statistics. If the summaries of the observed
data set fall within an acceptable range of the summaries of the simulated
data sets, then the model is deemed adequate.

We have already made many draws from the posterior predictive
distribution from
the analysis in Section~\ref{sec:weather-generator}, so we can use
those for
model assessment. Since we are interested in heat waves, we choose summary
statistics that describe the extremes of the temperature distribution. To
assess the marginal fit, we compute the 0.99 and 0.999 empirical quantiles.
To assess the fit of the dependence, we compute an estimate of the extremal
index [\citet{ferro-2003a}] at the 0.975 observed quantile
(denoted as
$\hat{\vartheta}$ in Table~\ref{tab:bayes-intervals}), which can be
interpreted as the inverse of the mean size of clusters of observations above
the chosen threshold, as well as $\hat{\chi}(u)$ at time lags of 1
and 5 days
and at several values of $u$. We compute each summary statistic for each
posterior predictive draw and report the 0.025 and 0.975 quantiles. If the
observed statistics fall within their corresponding predictive
intervals, we
declare that the model fits the data satisfactorily.

%
\begin{table}[b]
\tabcolsep=0pt
\caption{Posterior predictive intervals for summary statistics. Each column
corresponds to a summary statistic. The statistics $q_{0.99}$ and
$q_{0.999}$ (the empirical 0.99 and 0.999 quantiles) describe the
extremes of the marginal predictive distributions, and the
statistics $\hat{\vartheta}$ (the extremal index) and
$\hat{\chi}_h(u)$ (extremal dependence at time lag $h$ and
threshold $u$) describe the strength of the asymptotic dependence.
For both Paris and Moscow, the top and bottom rows correspond to
the 0.025 and 0.975 quantiles of the posterior draws, and the
middle row corresponds to the observed quantities}
\label{tab:bayes-intervals}
\begin{tabular*}{\tablewidth}{@{\extracolsep{\fill}}@{}lcccccccccc@{}}
\hline
& & $\bolds{q_{0.99}}$ & $\bolds{q_{0.999}}$ & $\bolds{\hat{\vartheta}}$ & $\bolds{\hat{\chi}_1(28)}$
& $\bolds{\hat{\chi}_5(28)}$ & $\bolds{\hat{\chi}_1(32)}$ & $\bolds{\hat{\chi}_5(32)}$
& $\bolds{\hat{\chi}_1(36)}$ & $\bolds{\hat{\chi}_5(36)}$ \\
\hline
& $q_{0.025}$ & 33.62 & 33.62 & 0.205 & 0.529 & 0.232 & 0.303 & 0.030 & 0.000 & 0.000 \\
Paris & obs. & 35.33 & 38.48 & 0.561 & 0.607 & 0.297 & 0.389 & 0.102 & 0.462 & 0.231 \\
& $q_{0.975}$ & 38.32 & 43.17 & 0.867 & 1.000 & 1.000 & 1.000 & 1.000 & 0.686 & 0.386
\\[3pt]
& $q_{0.025}$ & 32.50 & 34.83 & 0.225 & 0.494 & 0.165 & 0.244 & 0.000 & 0.000 & 0.000 \\
Moscow & obs. & 34.13 & 36.89 & 0.245 & 0.662 & 0.356 & 0.600 & 0.280 & 0.375 & 0.125 \\
& $q_{0.975}$ & 35.82 & 39.25 & 0.754 & 0.758 & 0.517 & 0.600 & 0.333 & 0.571 & 0.167 \\
\hline
\end{tabular*}
\end{table}

Table~\ref{tab:bayes-intervals} shows the results of the posterior predictive
checks. For both Paris and Moscow, the observed summary statistics fall
within the posterior 95\% intervals of the chosen statistics. The only
hint of a problem is for $\hat{\chi}(32)$ at lag 1 for Moscow, where the
observed statistic falls on the endpoint of the predictive interval. Other
than that, the predictive diagnostic indicates that the model provides a
suitable fit to the data.

\section{Discussion}

We have presented a simple Bayesian latent state model for studying heat
waves. The chief virtue of this model is its
interpretability; the latent state vector $\bS$ directly indicates
which days
are part of heat waves, and the Markov transition probabilities
represent the
frequency and duration of heat waves. In addition to the easy
interpretability, the latent state model has the advantage that, unlike other
extreme value models, there is no need to prespecify a threshold over which
the GPD applies or to use censored likelihoods for data below the threshold,
giving it a certain elegance.

One feature that is extremely useful is the ability to sample from the
fitted model, allowing it to act as a weather generator
(Section~\ref{sec:weather-generator}). In this way, our model is
adaptable to any
operational definition of heat waves that is germane to the application at
hand. An additional feature of
our approach is that missing values, endemic to meteorological data, are
easily handled by treating them as unknown parameters.

A limitation of our approach is that it only considers maximum daily
temperatures, but other aspects of heat waves might be of interest to
analysts. A more complete picture of heat waves might include, for example,
the daily minimum temperature or measures of heat stress.

In the case studies, we have assumed stationary that might not be realistic
for other data sets, but extensions are straightforward. A more intricate
analysis would allow, for example, a seasonal trend in $u$ and $\mu$. In
addition, it would be a simple matter to include an inter-annual trend for
model parameters, which would allow for the investigation of long-term changes
in the behavior of heat waves.

Another improvement would be to
borrow strength across space using Gaussian process priors, which would
provide improved statistical efficiency at the expense of a more complicated
forward-filtering backward-sampling algorithm. Candidates for spatial priors
are the Markov transition probabilities $a_0$ and $a_1$, the GPD parameters
$u, \sigma$ and $\xi$, and possibly even the dependence parameters
$\alpha,
\alpha_{01}$ and $\phi$. A related extension is to replace the two-state
Markov chain on the states with a logistic regression-type model, where the
state probabilities would depend on a spatially and temporally dependent
random process, and possibly on covariates as well. This type of model is
appealing, but it sacrifices much of the interpretability that is so
desirable.


\section*{Acknowledgments}
We would like to thank the three anonymous referees, as well as the Associate
Editor, for their insightful comments.

\begin{supplement}[id=suppA]
\stitle{Additional analysis}
\slink[doi]{10.1214/00-10.1214/15-AOAS873SUPP} 
\sdatatype{.pdf}
\sfilename{aoas873\_supp.pdf}
\sdescription{The Supplement contains additional exploratory analysis related
to the case study, a sensitivity analysis for the prior
distribution of the GPD threshold $u$, and results of the model
run on 2003 temperatures at several additional sites throughout
Western Europe.}
\end{supplement}

%

\printaddresses
\end{document}